\documentclass[aip,apl,amsmath,amssymb,reprint]{revtex4-1}
\usepackage{epsfig}
\usepackage{color}
\usepackage{graphicx}
\usepackage{dcolumn}
\usepackage{bm}
\usepackage{hyperref}

\begin{document}

\title{Clogging by sieving in microchannels: Application to the detection of contaminants in colloidal suspensions}

\author{Alban Sauret}
\affiliation{\mbox{Department of Mechanical and Aerospace Engineering, Princeton University, Princeton, New Jersey 08544, USA}} 

\author{Erin C. Barney}
\affiliation{\mbox{Department of Engineering, Trinity College, Hartford, Connecticut 06106, USA}} 

\author{Adeline Perro}
\affiliation{\mbox{Universit\'e de Bordeaux, CNRS, ISM UMR 5255, 33405 Talence Cedex, France}}

\author{Emmanuel Villermaux}
\altaffiliation[also at ]{Institut Universitaire de France, Paris}
\affiliation{\mbox{Aix Marseille Universit\'e, CNRS, Centrale Marseille, IRPHE UMR 7342, 13384 Marseille, France}}

\author{Howard A. Stone}
\affiliation{\mbox{Department of Mechanical and Aerospace Engineering, Princeton University, Princeton, New Jersey 08544, USA}}

\author{Emilie Dressaire} \email[Electronic mail: ]{dressaire@nyu.edu}
\affiliation{\mbox{Department of Mechanical and Aerospace Engineering, Princeton University, Princeton, New Jersey 08544, USA}}
\affiliation{\mbox{Department of Engineering, Trinity College, Hartford, Connecticut 06106, USA}} 
\affiliation{Department of Mechanical and Aerospace Engineering, New York University Polytechnic School of Engineering, Brooklyn, New York 11201, USA}

\date{18 july 2014}

\begin{abstract}
We report on a microfluidic method that allows measurement of a small concentration of large contaminants in suspensions of solid micrometer-scale particles. To perform the measurement, we flow the colloidal suspension through a series of constrictions, i.e. a microchannel of varying cross-section. We show and quantify the role of large contaminants in the formation of clogs at a constriction and the growth of the resulting filter cake. By measuring the time interval between two clogging events in an array of parallel microchannels, we are able to estimate the concentration of contaminants whose size is selected by the geometry of the microfluidic device. This technique for characterizing colloidal suspensions offers a versatile and rapid tool to explore the role of contaminants on the properties of the suspensions.
 \end{abstract}

\maketitle

\label{firstpage}


Particle-laden flows are ubiquitous in many common industrial applications such as the purification of water or oil by removing solid particles suspended in the fluid. Filtration methods generally rely on the capture of solid particles in a porous media or a filter. However, the efficiency of filters is reduced by clogging, which happens when particles lodge in the cross-section of the pore forming a blockage that prevents particles from flowing downstream; as a result, the flow rate through the system is reduced dramatically. Pioneering work on filtration characterized the efficiency of a filter through macroscopic measurements and theoretical models. \cite{ruth1933,ruth1935,carman1938,ruth1946,coulson1968,duclos2006,giglia2012,griffiths2014} More recently, the use of microfluidic methods allows the investigation at the pore-scale level \cite{wyss2006} of both filtration processes and model biological systems where hard or soft particles are suspended in complex fluids.\cite{chesnutt2009,cohen2013,altshuler2013,drescher2013} 

Clogging can also be a technological challenge for applications that require transport of colloidal suspensions through micro or milli-channels. Blockage of channels results in failure of devices such as inkjet printers or microfluidic systems where particles can be intentionally present or introduced by the surrounding environment in the form of dust or contaminants. 

The simplest clogging mechanism is based on steric effects: a particle larger than the pore blocks the inlet of the channel.\cite{han2007} Other mechanisms of clogging are also possible such as aggregation of particles against the wall of the channel  \cite{wyss2006,datta1998,ramachandran1999,sharp2005,bessiere2008,kampel2008,emilie,mustin2010,gudipaty2010,bacchin2011,agbangla2012,agbangla2014} or the jamming of concentrated suspension.\cite{goldsztein2004,campbell2010,conrad2010,genovese2011,holloway2011} In addition clogging can be used to measure the mechanical properties of flexible particles such as biological cells. \cite{guo2011,mcfaul2012,preira2013,amy2014} In order to prevent clogging or aggregation of particles\cite{gudipaty2010} and understand the mechanisms involved, it is critical to know the physicochemical properties of the colloidal suspension. 

In this work, we make use of dilute suspensions of rigid colloidal particles coated with hydrophilic groups, which limit the aggregation between particles and the PDMS (polydimethylsiloxane) walls so that clogging is due to steric effects alone. Relying on steric exclusion allows us to determine the size and concentration of contaminants, both simply and rapidly. The present method is more effective than traditional techniques to detect dilute contaminants whose size is several times greater than the mean diameter of the colloidal particles in suspension. To obtain an accurate estimate, optical measurements require several iterations making them time consuming. Recently, microfluidic methods have been developed  to sort particles using inertial flows: inertial effects lead to the alignment of particles of similar sizes and/or shapes \cite{dicarlo2007,dicarlo2009,dicarlo2009_review,dicarlo2014}. In the method reported here, inertial effects are not required as steric exclusion occurs even at low Reynolds numbers: we take advantage of the clogging of a microfluidic device to estimate the concentration of large particles, i.e. contaminants, in a colloidal suspension. 

\begin{center}
\begin{figure}
\includegraphics{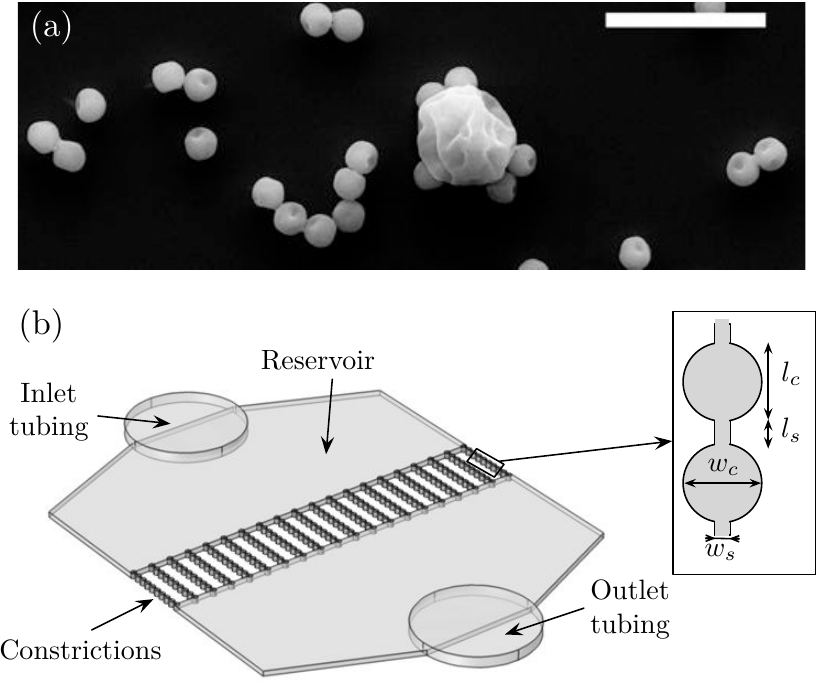}
\caption{(a) Scanning electron microscope (SEM) image of the colloidal suspension used in this study showing the presence of large contaminants among the $2 \, \mu{\rm m}$ diameter particles (scale bar is $10\,\mu{\rm m}$). (b) Schematic of the experimental setup. The colloidal suspension is injected at constant pressure $\Delta p$ into the reservoir, flows through the parallel microchannels and out of the device through a second reservoir. Inset: each microchannel is a series of large ($w_c={\ell}_c=50 \, \mu{\rm m}$) and small constrictions ($w_s=10 \, \mu{\rm m}$ and ${\ell}_s=20 \, \mu{\rm m}$). The height of the device is $h=14.2\,\mu{\rm m}$.}
\label{Figure1}
\end{figure}
\end{center}

In most studies, the suspensions are assumed to be monodisperse or at least to have a narrow distribution around the mean diameter of the particles. In practice, commercial suspensions often contain a small density of large contaminants. For instance, figure \ref{Figure1}(a) illustrates the presence of contaminants in a commercial colloidal suspension produced by the classical emulsion polymerization method.

In this Letter, we observe that the formation of clogs is determined solely by the number of large contaminants that flows through the device. We report a method to evaluate a small concentration of these contaminants of large size in a colloidal suspension. To achieve this goal, we flow the suspension in a microfluidic device consisting of an inlet reservoir followed by an array of parallel microchannels in which the particles of small size flow through while the contaminants clog the channels due to sieving. By measuring the times at which the clogging events occur, we directly estimate the concentration of contaminants of a size set by the geometry of the device.

In all of the experiments presented here, we use colloidal suspensions of polystyrene microspheres of mean diameter $2.12\,\mu{\rm m}$ (Polysciences, Inc.). The particles are coated by carboxylate groups to avoid the adsorption onto the PDMS walls of the microchannel and the formation of clusters of particles in suspension. We typically work with small solid volume fractions in the range $\phi=[10^{-4};\,10^{-2}]$ v/v such that the clogging events are not induced by jamming at the bottlenecks of the constrictions within the timescale of our experiments. 

The PDMS microfluidic device is made using standard soft lithography methods \cite{pdms1,pdms2} and is bonded using plasma treatment 24 hours prior to the experiments. An inlet tubing of radius $a_{\rm tub}=190 \,\mu{\rm m}$ brings the colloidal suspension into the microfluidic device through a large reservoir (width $w_{\rm res}=3600 \,\mu{\rm m}$, length ${\ell}_{\rm res}=1800 \,\mu{\rm m}$) as shown in figure \ref{Figure1}(b). The inlet reservoir feeds the suspension into $N=20$ parallel microchannels of smallest width $w_s=10\,\mu{\rm m}$ and largest width $w_c=50\,\mu{\rm m}$ as illustrated in the inset of figure \ref{Figure1}(b).  The unusual shape of the microchannel allows for non-spherical particles to reorient and clog the microchannel as we shall see later. The colloidal suspension flows through an outlet reservoir to exit the device. The height of the device is constant and equal to $h = 14.2 \,\mu{\rm m}$.

 The fluid is pushed through the device by imposing a constant pressure difference with a regulator (Omega AR91-005) in the range $\Delta p \in [1.7;34.5]\,{\rm kPa}$ such that the flow rate in each open microchannel or pore remains approximately constant over time. Because our device is  made of PDMS, the pressure difference induces deformation of the wide inlet reservoir.\cite{gervais2006,hardy2009} Therefore, the contaminants that contribute to the sieving process (diameter $D < 14.2 \,\mu{\rm m}$) are not confined and the velocity of the contaminants in the reservoir can be approximated as the velocity of the particles.\cite{S1}  In the vicinity of the parallel microchannels, however, the deformation is negligible because their typical width is much smaller ($w \leq 50\,\mu{\rm m}$). Therefore, only contaminants of size $D < h$ can flow into the constriction and contaminants of diameter $ D>w_s$ clog the channel by sieving. We thus select contaminants whose sizes are in the range $D \in [w_s;h]$. The flow rate of the suspension through the microfluidic device can be estimated using the definition of the hydraulic resistance of the device $R_{\rm h}$.\cite{bruus2008,oh2012} The pressure difference $\Delta p$ and the flow rate $Q$ satisfy the relation $\Delta p=R_{\rm h}\,Q$. The clogging of the microchannels is observed using a Leica DMI4000B inverted microscope and a Leica DFC360FX camera. From the movies, we acquire the time interval between two clogging events to obtain statistical information about the clogging process.

\begin{center}
\begin{figure}
\begin{center}\includegraphics{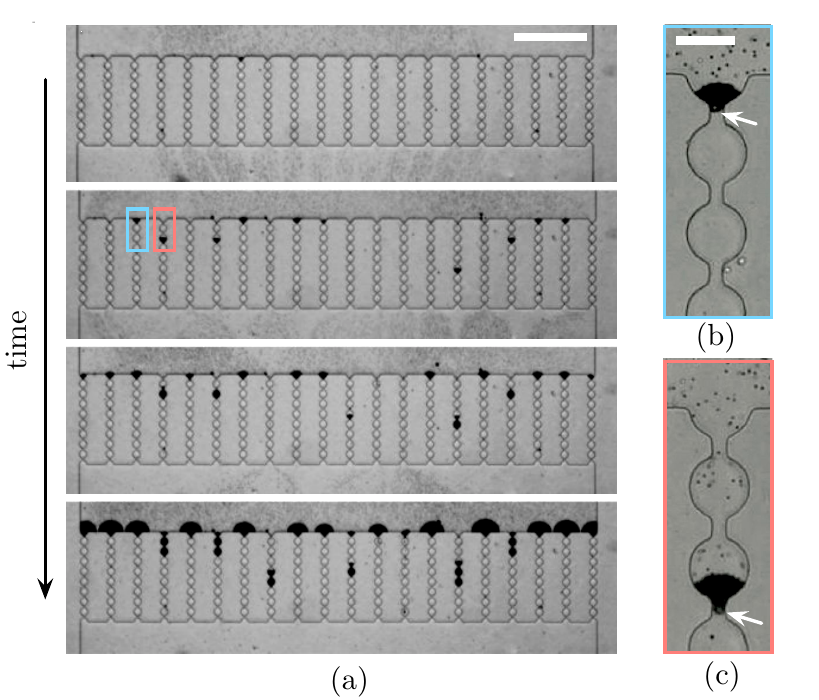}\end{center}
\caption{(a) Time lapse of a typical clogging experiment ($\Delta p=13.8$ kPa) in a device consisting of 20 parallel microchannels. Particles are black. The panels show the evolution of the number of clogged channels by sieving. After clogging, the particles accumulate and form a filter cake. The suspension ($\phi=2\times10^{-3}$) flows from top to bottom (scale bar is $500\, \mu{\rm m}$). Successive pictures are taken at $t=2$ s, $13$ s, $36$ s and $150$ s. (b) Close-up view of the clog formed when a large contaminant clogs a microchannel at the entrance or (c) further inside a microchannel. The white arrows show the position of the large contaminants and the scale bar is $50\, \mu{\rm m}$.}
\label{Figure2}
\end{figure}
\end{center}

During a typical experiment, clogs form successively blocking the channels until no pore remains open. An example of such a  clogging cascade is summarized by the time-lapse presented in figure \ref{Figure2}(a) for $\Delta p=13.8$ kPa and a solid volume fraction $\phi=2\times10^{-3}$.\cite{S1} The black regions correspond to the presence of aggregated particles whereas the lightest regions are the diluted suspension or pure water. Initially, the colloidal suspension flows through the device and no clog is observed: all of the microchannels are open (top image). When a large contaminant arrives at a constriction, it clogs the channel because of steric effects. After clogging, the smaller particles accumulate against the immobile contaminant, which leads to the formation of a filter cake (in black in the pictures).\cite{ortiz2013,ortizsubmitted} A close-up view of a filter cake demonstrates that the clog is initially generated by a large particle, i.e. a contaminant (see figure \ref{Figure2}(b)). We note that some clogs are not formed at the entrance of the channel but rather within the constricted microchannel (figure \ref{Figure2}(c)). This behavior could be a consequence of the shape of the contaminants that can be non-spherical. For instance, we observe contaminants with an elongated ellipsoidal shape.\cite{S1} Such an anisotropic particle is able to reorient a few times before clogging the channel, as observed with fibers.\cite{berthet2013} One by one the channels clog and we record the clogging cascade. We analyze the images with a custom-written
MATLAB routine, determine the clogging time for each channel and extract the distribution of the clogging time intervals between two channels.

\begin{center}
\begin{figure}
\begin{center}\includegraphics{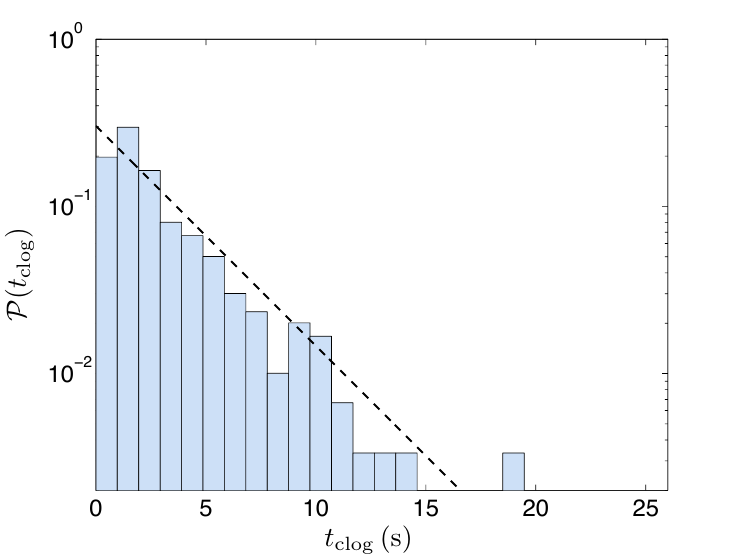}\end{center}
\caption{Distribution of clogging time intervals $t_{\rm clog}$ for $\Delta p=13.8$ kPa and $\phi=2\times10^{-3}$. The black dotted line is the best fit assuming a Poisson distribution with $\langle t_{\rm clog} \rangle =3.3$ s. }
\label{Figure3}
\end{figure}
\end{center}


From the distribution of the clogging time intervals (see figure \ref{Figure3}), we can estimate the concentration of contaminants in the colloidal suspension. Indeed, the flow rate when the $i$-th channel clogs is $Q(i)=\Delta p/R_{\rm h}(i)$ with $R_{\rm h}(i) \simeq 2\,R_{\rm tub}+2\,R_{\rm res}+R/(N-i)$. In this expression $R_{\rm res}$, $R_{\rm tub}$ and $R$ denotes the hydraulic resistance of reservoir, tubing and one non-clogged microchannel, respectively. We assume that once a microchannel is clogged, its flow rate becomes zero and its hydraulic resistance $R \to + \infty$. The hydraulic resistance depends on the geometrical properties of the channel and can be expressed analytically for a rigid microfluidic device. \cite{bruus2008,oh2012} However, the geometry of the constricted channel as well as the deformation of PDMS\cite{gervais2006,hardy2009} make it difficult to estimate theoretically the hydraulic resistance of the device. Therefore we rely on an experimental measurement of the hydraulic resistance of the different parts of the microfluidic device. We find that in our system $2\,R_{\rm tub}+2\,R_{\rm res}=(7.92 \pm 0.57)\times 10^{12}\,\rm{kg\,m^{-4}\,s^{-1}}$ and $R=(9.15 \pm 0.14)\times 10^{13} \,\rm{kg\,m^{-4}\,s^{-1}}$.

We define $c$, the concentration of large contaminants susceptible to forming a clog in a channel. The clogging time interval between the clogging of the $i$-th and the $(i-1)$-th channel is expressed as $ t_{\rm clog,i}=t_{\rm i}-t_{\rm i-1}={1}/{[c\,Q(i)]}$. Therefore the mean time interval $\langle t_{\rm clog} \rangle$ between two clogging events is given by
\begin{equation}\label{tclog}
\langle t_{\rm clog} \rangle =\frac{1}{N}\sum_{i=0}^{N-1}\frac{2\,R_{\rm tub}+2\,R_{\rm res}+{R/(N-i)}}{c\,\Delta p}.
\end{equation}
The mean clogging time interval is determined experimentally. The time interval distribution is well fitted by a Poisson distribution
\begin{equation}
\mathcal{P}( t_{\rm clog})=\frac{1}{\langle t_{\rm clog} \rangle}\,\textrm{exp}\left(-\frac{t_{\rm clog}}{\langle t_{\rm clog} \rangle}\right)
\end{equation}
\noindent with $\langle t_{\rm clog} \rangle =3.3\pm 0.5$ s for $N=20$. Using Equation (\ref{tclog}), we obtain that the concentration of the large contaminants of size in the range $D \in [10;14.2]\;\mu$m is $c \simeq (5.6\,\pm\,1.1)\,\times 10^{8}\,{\rm m^{-3}}$ for the $2\times10^{-3}$ v/v suspension. We can thus determine the relative concentration of large contaminants in the colloidal suspension, i.e. the ratio $f_{\rm c}$ of the concentration of contaminants $c$ to that of colloidal particles $c_{\rm part}$: $f_{\rm c}=c/c_{\rm part}=1.4\pm0.3\,\times 10^{-6}$.

To ensure that our method is reliable, we also estimate the concentration of large contaminants using direct visualization. The obtained value is of the order of $f_{\rm c} \simeq 10^{-6} - 10^{-5}$ and is in fairly good agreement with our microfluidic method regarding the precision on the determination of the size of the contaminants using the direct visualization method. Indeed, we should note that the uncertainties on the range of size of the contaminant is much larger using direct visualization.

\begin{center}
\begin{figure}
\begin{center}\includegraphics{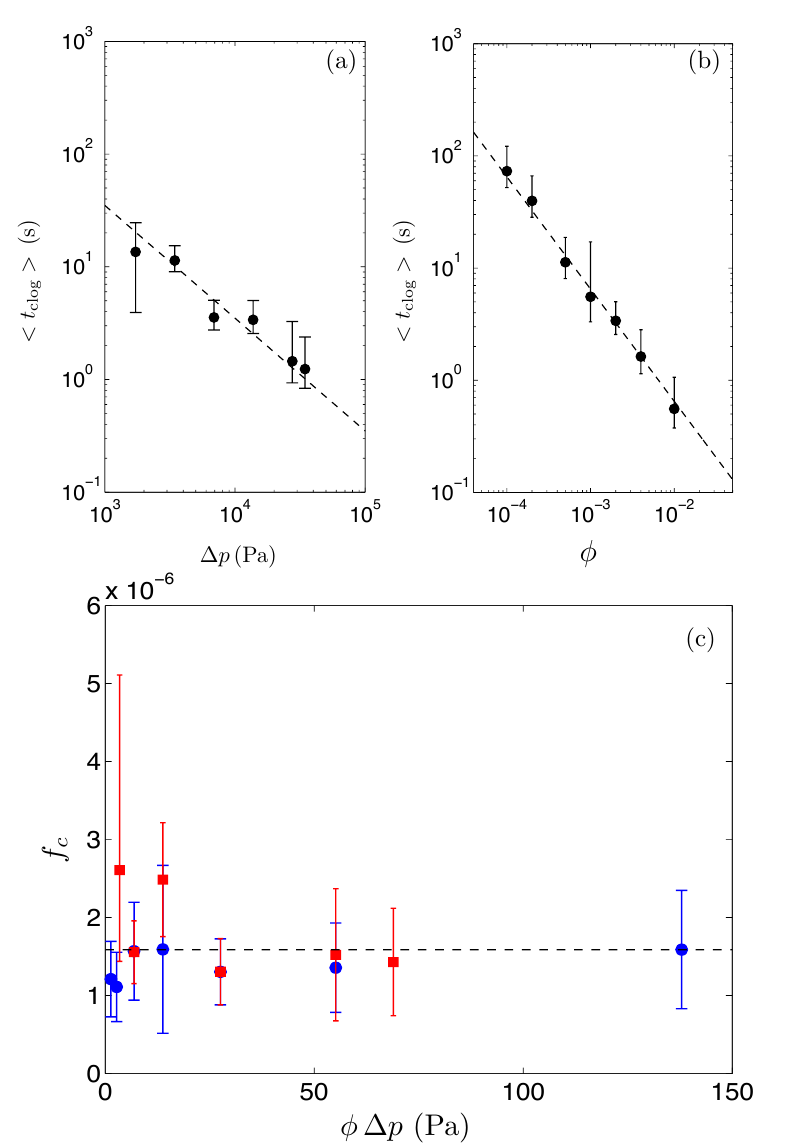}\end{center}
\caption{(a) Evolution of the mean clogging time as a function of the pressure difference across the device for $\phi=2\times10^{-3}$. The dashed line has a slope proportional to $1/\Delta p$. (b) Evolution of the mean clogging time as a function of the particle volume fraction for  $\Delta p=13.8$ kPa where the dashed line has a slope proportional to $1/\phi$. (c) Relative concentration of large contaminants in the colloidal suspension $f_{\rm c}$ as a function of $\Delta p\,\phi$ calculated using relation (\ref{solution_fc}). The blue circles correspond to $\Delta p=13.8$ kPa and $\phi\in[10^{-4};10^{-2}]$, the red squares correspond to $\phi=2\times10^{-3}$ and $\Delta p \in [1.7;34.5]\,{\rm kPa}$. The horizontal dashed line represents the mean value, $f_{\rm c}=1.6\,\times 10^{-6}$.}
\label{Figure4}
\end{figure}
\end{center}

Our experimental approach is a robust method to determine the concentration of contaminants in colloidal suspensions. We explore the influence of the control parameters: the imposed pressure difference and the concentration of the colloidal suspension. In order to study the role of those parameters, all of the suspensions are prepared from a single mother solution, with a fixed contaminant concentration. The mean clogging interval depends on the number of contaminants that enter the device per unit time. The total number of particles that flow through a microchannel is proportional to the volume of suspension and the solid volume fraction $\phi$. Therefore the mean clogging interval should scale as 
\begin{equation}
\langle t_{\rm clog} \rangle \propto \frac{1}{Q\,\phi} \propto \frac{1}{\Delta p\,\phi}.
\end{equation}
Both the predicted variation with the pressure difference (figure \ref{Figure4}(a)) and the solid volume fraction of the colloidal suspension (figure \ref{Figure4}(b)) are captured experimentally. In addition, using the results obtained for all of these experiments allows us to estimate the relative concentration $f_{\rm c}$ of large contaminants in the mother colloidal suspension with a better accuracy. We can estimate $f_{\rm c}=c/c_{\rm part}$ using relation (\ref{tclog})
\begin{equation}\label{solution_fc}
f_{\rm c}=\frac{4\,\pi\,r^3}{3\,\Delta p\,\phi\,N \langle t_{\rm clog} \rangle }\sum_{i=0}^{N-1}\left(2\,R_{\rm tub}+2\,R_{\rm res}+\frac{R}{N-i} \right)
\end{equation}
where $r \simeq 1.06\,\mu{\rm m}$ is the radius of the small particles. The obtained value of $f_{\rm c}$ for varying pressure difference $\Delta p$ and concentration of the suspension $\phi$ is constant and equal to $f_{\rm c}=1.6\,\times 10^{-6}$ as illustrated in figure \ref{Figure4}(c). This result is in agreement with the value obtained previously for $\Delta p=13.8$ kPa and $\phi=2\times10^{-3}$. It confirms that for a given geometry of the device, i.e. a given hydraulic resistance, which can be determined experimentally, the measurement of the mean clogging time interval $\langle t_{\rm clog} \rangle $ leads to a good estimate of the concentration of large contaminants in the colloidal suspension. In addition, the experiments can be performed for a broad range of pressure differences or suspension concentrations, which allows for the tuning of the experimental parameters to the concentration of large contaminants. For a large concentration of contaminants, one may want to use a very dilute suspension and/or a small pressure difference to avoid very short experiments, whereas for a small concentration of contaminants, a concentrated suspension and/or a large pressure difference is preferred. 

In this Letter, we report an approach to evaluate minute amounts of contaminants in a colloidal suspension. The concentration is determined by measuring the time distribution of the clogging events in an array of microchannels. Our approach is simple and rapid to implement as it only requires a few minutes to run tests with low-cost microfluidic devices. We demonstrate how the clogging interval relates to the concentration of large contaminants in the suspension. By controlling the height of the channel and the minimum width of the constriction we are able to detect contaminants of diameter $w_s<d<h$.  The agreement between our measurements and direct observations is an important validation of our method. This work demonstrates and takes advantage of the importance of large contaminants in confined particle-laden flows, a topic that remains largely unexplored.  
\vspace{0.5cm}

We gratefully acknowledge the NSF for support via grant CBET-1234500 to HAS and the CT Space Grant Pro-Sum No. 649 to ECB and ED.


\begin{thebibliography}{0}%
\makeatletter
\providecommand \@ifxundefined [1]{%
 \@ifx{#1\undefined}
}%
\providecommand \@ifnum [1]{%
 \ifnum #1\expandafter \@firstoftwo
 \else \expandafter \@secondoftwo
 \fi
}%
\providecommand \@ifx [1]{%
 \ifx #1\expandafter \@firstoftwo
 \else \expandafter \@secondoftwo
 \fi
}%
\providecommand \natexlab [1]{#1}%
\providecommand \enquote  [1]{``#1''}%
\providecommand \bibnamefont  [1]{#1}%
\providecommand \bibfnamefont [1]{#1}%
\providecommand \citenamefont [1]{#1}%
\providecommand \href@noop [0]{\@secondoftwo}%
\providecommand \href [0]{\begingroup \@sanitize@url \@href}%
\providecommand \@href[1]{\@@startlink{#1}\@@href}%
\providecommand \@@href[1]{\endgroup#1\@@endlink}%
\providecommand \@sanitize@url [0]{\catcode `\\12\catcode `\$12\catcode
  `\&12\catcode `\#12\catcode `\^12\catcode `\_12\catcode `\%12\relax}%
\providecommand \@@startlink[1]{}%
\providecommand \@@endlink[0]{}%
\providecommand \url  [0]{\begingroup\@sanitize@url \@url }%
\providecommand \@url [1]{\endgroup\@href {#1}{\urlprefix }}%
\providecommand \urlprefix  [0]{URL }%
\providecommand \Eprint [0]{\href }%
\providecommand \doibase [0]{http://dx.doi.org/}%
\providecommand \selectlanguage [0]{\@gobble}%
\providecommand \bibinfo  [0]{\@secondoftwo}%
\providecommand \bibfield  [0]{\@secondoftwo}%
\providecommand \translation [1]{[#1]}%
\providecommand \BibitemOpen [0]{}%
\providecommand \bibitemStop [0]{}%
\providecommand \bibitemNoStop [0]{.\EOS\space}%
\providecommand \EOS [0]{\spacefactor3000\relax}%
\providecommand \BibitemShut  [1]{\csname bibitem#1\endcsname}%
\let\auto@bib@innerbib\@empty
\end{thebibliography}%


\begin{thebibliography}{1}


\bibitem{ruth1933}
B. F. Ruth, G. H. Montillon, and R. E. Montonna, Ind. Eng. Chem., {\textbf{25}}, 76-82 (1933).

\bibitem{ruth1935}
B. F. Ruth, Ind. Eng. Chem., {\textbf{27}}, 708-723 (1935).

\bibitem{carman1938}
P. C. Carman, Trans. Inst. Chem. Eng, {\textbf{16}}, 168-88 (1938).

\bibitem{ruth1946}
B. F. Ruth, Ind. Eng. Chem., {\textbf{38}}, 564-571 (1946).

\bibitem{coulson1968}
J. M. Coulson \& J. F. Richardson, {\textit{Chemical Engineering}}. New York: Pergamon (1968).

\bibitem{duclos2006}
C. Duclos-Orsello, W. Li, and C. C. Ho, J. Membrane Sci. {\textbf{280}}, 856Ð866 (2006).

\bibitem{giglia2012}
S. Giglia, and G. Straeffer, J. Membrane Sci. {\textbf{417-418}}, 144Ð153 (2012).

\bibitem{griffiths2014}
I. M. Griffiths, A. Kumar, and P. S. Stewart, J. Membr. Sci. {\textbf{432}}, 10-18 (2014).

\bibitem{wyss2006}
H. M. Wyss, D. L. Blair, J. F. Morris, H. A. Stone, and D. A. Weitz, Phys. Rev. E {\textbf{74}}, 061402 (2006).


\bibitem{cohen2013}
S. Cohen, and L. Mahadevan, Phys. Rev. Lett. {\textbf{110}}, 138104 (2013).

\bibitem{chesnutt2009}
J. K. W. Chesnutt, and J. S. Marshall, Microvasc. Res. {\textbf{78}}, 301--313 (2009).

\bibitem{altshuler2013}
E. Altshuler, G. Mi{\~n}o, C. P{\'e}rez-Penichet, L. del R{\'\i}o, A. Lindner, A. Rousselet, and E. Cl{\'e}ment Soft Matter {\textbf{9}}, 1864 (2013).

\bibitem{drescher2013}
K. Drescher, Y. Shen, B. L. Bassler, and H. A. Stone, Proc. Natl. Acad. Sci. {\textbf{110}}, 4345 (2013)





\bibitem{han2007}
J. Han, J. Fu, and R. B. Schoch, Lab Chip {\textbf{8}}, 23 (2008).

\bibitem{datta1998}
S. Datta, and S. Redner, Phys. Rev. E {\textbf{58}}, R1203 (1998).

\bibitem{ramachandran1999}
V. Ramachandran, and H. S. Fogler, J. Fluid Mech. {\textbf{385}}, 129-156 (1999). 

\bibitem{sharp2005}
K. V. Sharp, and R. J. Adrian, Microfluid. Nanofluid. {\textbf{1}}, 376-380 (2005).


\bibitem{bessiere2008}
Y. Bessiere, D. F. Fletcher, and P. Bacchin, J. Membr. Sci. {\textbf{313}}, 52--59 (2008).

\bibitem{kampel2008}
G. Kampel, G. H. Goldsztein, and J. C. Santamarina, App. Phys. Lett. {\textbf{92}}, 084101 (2008).

\bibitem{emilie}
E. Dressaire, {\textit{Shaping Fluid-Fluid Interfaces: From Molecular Monolayers to Thin Liquid Films}}. PhD dissertation, Harvard University (2009).

\bibitem{mustin2010}
B. Mustin, and B. Stoeber, Microfluid. Nanofluid. {\textbf{9}}, 905-913 (2010). 

\bibitem{gudipaty2010}
T. Gudipaty, M. T. Stamm, L. S. Cheung, L. Jiang, and Y. Zohar, Microfluid. Nanofluid. {\textbf{10}}, 661--669 (2010).

\bibitem{bacchin2011}
P. Bacchin, A. Marty, P. Duru, M. Meireles, and P. Aimar, Adv. Colloid Interface Sci. {\textbf{164}}, 2-11 (2011). 

\bibitem{agbangla2012}
G.C. Agbangla, \'E. Climent, and P. Bacchin, Sep. Purif. Technol. {\textbf{101}}, 42-48 (2012).

\bibitem{agbangla2014}
G. C. Agbangla, P. Bacchin, and E. Climent, Soft Matter {\textbf{10}}, 6303-6315 (2014).


\bibitem{goldsztein2004}
G. H. Goldsztein, and J. C. Santamarina, Appl. Phys. Lett. {\textbf{85}}, 4535 (2004).



\bibitem{campbell2010}
A. I. Campbell, and M. D. Haw, Soft Matter {\textbf{6}}, 4688 (2010).

\bibitem{conrad2010}
J. C. Conrad, and J. A. Lewis, Langmuir {\textbf{26}}, 6102--6107 (2010).

\bibitem{genovese2011}
D. Genovese, and J. Sprakel, Soft Matter {\textbf{7}}, 3889 (2011).

\bibitem{holloway2011}
W. Holloway, J. M. Aristoff, and H. A. Stone, Phys. Fluids {\textbf{23}}(8), 081701 (2011).   


\bibitem{guo2011}
Q. Guo, S. M. McFaul, and H. Ma, Phys. Rev. E {\textbf{83}}, 051910 (2011).

\bibitem{mcfaul2012}
S. M. McFaul, B. K. Lin, and H. Ma, Lab Chip {\textbf{12}}, 2369-2376 (2012).

\bibitem{preira2013}
P. Preira, V. Grandne, J. M. Forel, S. Gabriele, M. Camara, and O. Theodoly, Lab Chip {\textbf{13}}, 161-170 (2013).

\bibitem{amy2014}
D. J . Hoelzle, C. K. Chan, B. A. Varghese, and A. C. Rowat, ÒA microflui-
dic technique to probe cell deformability,Ó J. Vis. Exp. (to be published).




\bibitem{dicarlo2007}
D. Di Carlo, D. Irimia, R. G. Tompkins, and M. Toner, Proc. Natl. Acad. Sci. {\textbf{104}}, 18892-18897 (2007).

\bibitem{dicarlo2009}
D. Di Carlo, J. Edd, K. Humphry, H. A. Stone, and M. Toner, Phys. Rev. Lett. {\textbf{102}}, 094503 (2009).

\bibitem{dicarlo2009_review}
D. Di Carlo, Lab Chip {\textbf{9}}, 3038 (2009).

\bibitem{dicarlo2014}
H. Amini, W. Lee, and D. Di Carlo, Lab Chip {\textbf{14}}, 2739 (2014).

\bibitem{pdms1}
Y. N. Xia and G. M. Whitesides, Annu. Rev. Mater. Sci. {\textbf{28}}, 153 (1998).

\bibitem{pdms2}
T. Squires and S. Quake, Rev. Mod. Phys. {\textbf{77}}, 977 (2005).

\bibitem{gervais2006}
T. Gervais, J. El-Ali, A. Gunther, and K. F. Jensen, Lab Chip {\textbf{6}}, 500-507 (2006).

\bibitem{hardy2009}
B. S. Hardy, K. Uechi, J. Zhen, and K. H. Pirouz, Lab Chip {\textbf{9}}, 935 (2009).

\bibitem{S1}
See supplementary material at [URL will be inserted by AIP] for corresponding movies.


\bibitem{bruus2008}
H. Bruus, {\textit{Theoretical Microfluidics}}. Oxford University Press (2008).

\bibitem{oh2012}
K.W. Oh, K. Lee, B. Ahn, and E.P. Furlani, Lab Chip {\textbf{12}}, 515-545 (2012). 


 
\bibitem{ortiz2013}
C. P. Ortiz, R. Riehn, and K. E. Daniels, Soft Matter {\textbf{9}}, 543-549 (2013). 

\bibitem{ortizsubmitted}
C. P. Ortiz, K. E. Daniels, and R. Riehn, ``Nonlinear elasticity of flow-stabilized solids,'' Phys. Rev. E (submitted).

\bibitem{berthet2013}
H. Berthet, M. Fermigier, and A. Lindner, Phys. Fluids {\textbf{25}}, 103601 (2013). 


\end{thebibliography}
\end{document}